# Paper-based ZnO self-powered sensors and nanogenerators by plasma technology

Xabier García-Casas,[1] Francisco J. Aparicio,[1,2]* Jorge Budagosky,[1,2]* Ali Ghaffarinejad,[1,§] Noel Orozco-Corrales,[1] Kostya (Ken) Ostrikov,[3] Juan R. Sánchez-Valencia,[1] Ángel Barranco[1] and Ana Borrás[1]*

[1]Nanotechnology on Surfaces and Plasma Lab, Materials Science Institute of Seville (CSIC-US), c/ Américo Vespucio 49, 41092, Seville, Spain.

[2]Departamento De Física Aplicada I, Escuela Politécnica Superior, Universidad De Sevilla, C/ Virgen De África 7, 41011, Seville, Spain.

[3]School of Chemistry and Physics and QUT Centre for Materials Science, Queensland University of Technology (QUT), Brisbane, QLD 4000, Australia.

[§]Current address: Sensors and Smart Systems Group, Institute of Engineering, Hanze University of Applied Sciences, 9747 AS Groningen, The Netherlands.

anaisabel.borras@icmse.csic.es; jorge.budagosky@icmse.csic.es; fjaparicio@icmse.csic.es

**ABSTRACT:** Nanogenerators and self-powered nanosensors have shown the potential to power low-consumption electronics and human-machine interfaces, but their practical implementation requires reliable, environmentally friendly and scalable, processes for manufacturing and processing. This article presents a plasma synthesis approach for the fabrication of piezoelectric nanogenerators (PENGs) and self-powered sensors on paper substrates. Polycrystalline ZnO nanocolumnar thin films are deposited by plasma-enhanced chemical vapour deposition on common paper supports using a microwave electron cyclotron resonance reactor working at room temperature yielding high growth rates and low structural and interfacial stresses. Applying Kinetic Monte Carlo simulation, we elucidate the basic shadowing mechanism behind the characteristic microstructure and porosity of the ZnO thin films, relating them to an enhanced piezoelectric response to periodic and random inputs. The piezoelectric devices are assembled by embedding the ZnO films in PMMA and using Au electrodes in two different configurations: laterally and vertically contacted devices. We present the response of the laterally connected devices as a force sensor for low-frequency events with different answers to the applied force depending on the impedance circuit, i.e. load values range, a behaviour that is theoretically analyzed. The vertical devices reach power densities as high 80 nW/cm2 with a mean power output of 20 nW/cm2. We analyze their actual-scenario performance by activation with a fan and handwriting. Overall, this work demonstrates the advantages of implementing plasma deposition for piezoelectric films to develop robust, flexible, stretchable, and enhanced-performance nanogenerators and self-powered piezoelectric sensors compatible with inexpensive and recyclable supports.

## Introduction

The growing demand for powering wireless devices has fostered the development of energy-harvesting solutions such as piezoelectric, pyroelectric, triboelectric, and thermoelectric nanogenerators.[1–3] These systems can convert the residual (kinetic or thermal) energy to electricity in the surrounding of a device to feed a small energy storage system or to be used as a primary power unit. However, the payback time of this technology is still non-competitive when compared with the use of batteries or against other energy conversion systems such as solar cells. Consequently, radically new approaches to optimize the performance and multifunctionality of nanogenerators while following the circular green economy principles, are crucial. As one of the most common materials, paper has attracted strong interest as a flexible substrate across several fields, from sensors and smart surfaces to electronics.[4–6] Owing to the latest advances in flexible electronics, cellulose paper's flexibility, and foldability have established it as a viable candidate for low-cost wearable electronic devices that can be integrated into curved surfaces. Major advantages of such implementation include lightweight, portability, disposability, recyclability, biodegradability, good printability, ease of fabrication, and breathability.[5,7,8] The high prospects of paper-based electronic devices arise from the ultralow material and manufacturing costs compared to other materials such as silicon and plastics. Also, low-cost electronic applications could be eventually achieved considering that the lifetime of electronic devices is becoming shorter and shorter.

Hence, paper-based electronics has experienced major advances in diverse fields[4,9] such as sensing,[10–12] energy harvesting[7,13,14] and storage,[15–17] and optoelectronics.[18] Its implementation into wearable electronic devices includes diverse types of skin-interfaced biomedical sensors[6,8,19–21] (temperature, strain, pressure, pH, biochemical composition and others) where cellulose paper plays a critical role as both flexible active material and/or substrate. A major category of those paper-based skin-interfaced sensors is biomechanical sensors providing real-time and continuous monitoring of either vital signals, such as respiration, pulse waveform or another biomechanical stimulus (e.g., acoustic ad).[8] Thus, within the context of the present Internet of Things (IoT) revolution, paper-based energy harvesters and sensors have high prospects for the development of disposable and wireless skin-interfaced biomedical devices that are monolithically integrated on the same substrate material. For this application, the power unit should have the same characteristics as the active device so the fully integrated paper system will be disposable, eco-friendly, and cost-effective.

Here we present a reliable process for the on-paper fabrication of thin film piezoelectric nanogenerators based on ZnO prepared by plasma enhanced chemical vapour deposition (PECVD). Plasma technology is considered an alternative sustainable approach in diverse fields ranging from agriculture and nanomedicine to chemical catalysis and advanced manufacturing of materials and devices.[22–24] Even though a current trend is plasma processing at atmospheric or close to atmospheric pressures, the fabrication of functional (low dimensional) materials and nanostructured surfaces is dominated by vacuum-based deposition methods. Specifically, techniques such as magnetron sputtering, plasma etching, reactive ion etching, and PECVD are widely exploited in optics and photonics, microelectronics, biomaterials, and energy.[25–28] Critical advantages such as low-temperature deposition when compared with pure vacuum approaches (CVD, ALD), high homogeneity, high yield, solventless character, and reduced or null production of residues make these techniques attractive in comparison with solvent-phase procedures. Such beneficial features have already opened the path for the implementation of plasma technologies in energy harvesting and self-powered nanosensors. Thus, examples can be found in a two-way transfer between one and the other topic. On one hand, the use of nanogenerators for the production of plasmas at atmospheric or near atmospheric pressure and the application of plasmas for the conditioning and improvement of triboelectric power management circuits. On the other hand, plasma-assisted processing and deposition techniques have been utilized in the fabrication of electrodes, piezoelectric layers and nanostructures, as well as the functionality and roughness optimization of triboelectric layers.

We have previously reported the formation of polycrystalline ZnO layers working as piezoelectric counterparts in core@multishell piezoelectric and hybrid piezo-triboelectric nanogenerators.[29,30] In those examples, thin ZnO shells ranging from tens to hundreds of nanometers were conformally synthesized on one-dimensional conductive ONWs@Ag or @Au electrodes and implemented on ITO/PET supports. Herein, we implement a step change by demonstrating the room temperature and low-cost fabrication of ZnO thin films for the development of paper-based piezoelectric nanogenerators and self-powered sensors (Scheme 1). At the fundamental level, we will apply Kinetic Monte Carlo simulations to gain insights into the crystalline and texture development of the thin films and also to reveal the role played by the substrate morphology and roughness in the growth of porous nanocolumnar arrangements. From the applied point of view, we will show a facile step-by-step procedure for the fabrication of robust piezoelectric nanogenerators on commercially available paper substrates and their application as real-scenario self-powered sensors. With these aims in mind, we have fabricated two types of architectures as shown in Scheme 1 (see also Figure S1 for devices photographs), namely laterally (a) and vertically (top-bottom) (b, cantilever) connected devices.

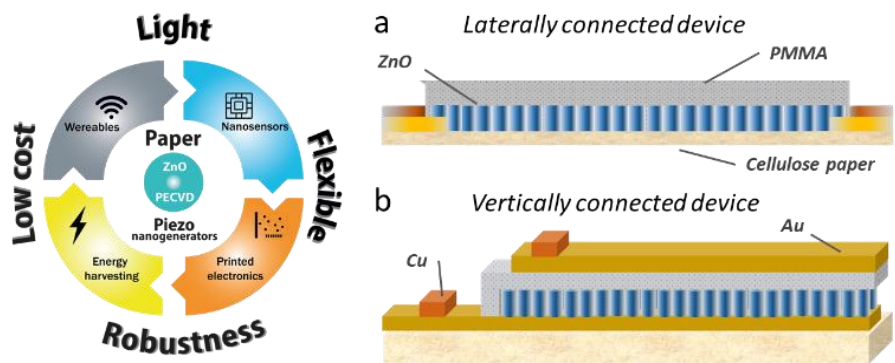

**Scheme 1. Device assembly.** Left) Conceptual map for paper-based energy harvesting and self-powered sensors. Right) Schematic on the layer-by-layer assembly for the laterally (a) and vertically (top-bottom) (b) connected devices. Please note, the scheme is not on a real scale.

## Experimental Methods

**Device fabrication and assembly**

ZnO thin films were fabricated by Plasma Enhanced Chemical Vapour Deposition (PECVD) at room temperature following the protocol detailed in the references [29–31]. Diethylzinc (($CH_2CH_3$)$_2$Zn) from Sigma-Aldrich was used as delivered as Zn precursor. The precursor was dosed to the reactor through a regulable valve and a shower-like dispenser positioned over the samples. The precursor was kept at RT meanwhile the tubes and dispenser were heated up to 80 ºC. Oxygen was used as plasma gas and inserted in the reactor by using a mass flow controller. The base pressure of the chamber was lower than $10^{-5}$ mbar, and the total pressure during the deposition was around $10^{-2}$ mbar. The partial pressure produced by the precursor during the reaction was $1.3\times10^{-3}$ mbar. The plasma was generated in a 2.45 GHz microwave Electron-Cyclotron Resonance (ECR) SLAN-II source operating at 500 W and the deposition was carried out in the downstream region of the reactor. The growth rate was settled at 10 nm/min. The thicknesses of the ZnO thin films were controlled by simply adjusting the deposition time to get the desired thickness after calibration by SEM on a Si(100) reference substrate.

Au electrodes were deposited by thermal evaporation in a vacuum chamber with a base pressure of $10^{-6}$ mbar. Substrates were cooled down at 20ºC during the evaporation and the deposition rate and thickness were controlled using a quartz crystal microbalance at 0.5 Å/s and 100 nm, respectively.

For the laterally connected devices (Scheme 1 a)), pieces of cellulose paper Vistacopy of 1 x 6 cm were sliced. First, Au electrodes were deposited by evaporation through a shadow mask, which avoids the deposition in the central area (ca. 1 x 2 cm) of the paper substrates (see Figure S1). The system was then coated with PMMA by spin coating. The PMMA polymer was dissolved in toluene 5% w/w and spun for 45 seconds at 1500 rpm and then heated up to 80ºC to remove the remaining solvent. The thickness of the PMMA layer was estimated by SEM on a Si(100) reference substrate at 600 nm.

The vertically connected systems (Scheme 1b)) were fabricated by first depositing the Au bottom electrode on the cellulose paper, leaving a small part of the paper substrate uncovered, which will allow making the final top electrical connection (see Figure S1). After, the ZnO thin film was grown, and then a PMMA layer was coated on top of the whole system and heated at 80ºC. This process was repeated three times consecutively. The thickness of the PMMA for vertically arranged devices was thicker than in the lateral one to avoid short-circuiting issues usually appearing on porous systems. Finally, the top Au electrode was deposited through a shadow mask with a P shape.

**Materials and device characterization**

SEM micrographs were acquired in a Hitachi S4800 working at 2 kV at working distances in the range of 2–4 mm. Samples were characterized as grown, without any additional conductive coating. The crystalline structure was analyzed by X-Ray Diffraction (XRD) spectrometer in a Panalytical X'PERT PRO model operating in the θ - 2θ configuration and using the Cu Kα (1.5418 Å) radiation as an excitation source. XRD data were also used to calculate the residual stress of ZnO coatings deposited on cellulose paper (coating thickness 5 μm) and of a thin film (500 nm) deposited on rigid fused silica. According to the biaxial stress model the residual in-plane stress $\sigma_c$ is calculated by:

$$\boldsymbol{\sigma_c = \left[\frac{2C_{13}^2 - C_{33}(C_{11}+C_{12})}{2C_{13}}\right] \times \varepsilon_{zz} = -233 \times \varepsilon_{zz} [GPa]} \quad \text{(Eq. 1)}$$

where $\varepsilon_{zz}$ is the strain in the lattice along the c-axis and $C_{ij}$ are the bulk elastic stiffness constants ($C_{11}$ = 208.8 GPa, $C_{12}$ = 119.7 GPa, $C_{13}$ = 104.2 GPa and $C_{33}$ = 213.8 GPa)**[32–35]**. In Eq. 1 the strain $\varepsilon_{zz}$ = ($d_{film}$-$d_0$)/$d_0$ is determined from the interplanar distance between planes (002) measured in the film by XRD and that of an unstrained bulk ZnO wurtzite structure ($d_0$ = 2.6033 Å)**[36]**. XRD peak position was determined by Voigt fitting after straight-line subtraction.

UV-VIS-NIR transmittance spectra were acquired in a PerkinElmer Lambda 750 UV/vis/NIR model. Room temperature water adsorption isotherms were obtained by the Quartz Crystal Microbalance (QCM) method developed by the authors.[37,38] The temperature of the microbalance was kept constant at 20-21 ºC. Before the acquisition of the water isotherm, the QCM deposited crystal was gently warmed under vacuum conditions ($10^{-6}$ mbar) up to 105 ºC to remove condensed water in the pores. The endpoint of this process was determined by the QCM signal. It must be noted that such characterization is herein only intended for qualitative comparison between the ZnO deposited on a completely flat substrate (QCM chip surface) and a roughness-modified substrate (i.e. Au particles layer deposited on top of the QCM crystal). The Au layer was fabricated following the same steps as for the bottom contact on the vertical paper devices.

Electrical measurements were carried out with a Keithley 2635A sourcemeter in combination with a Tektronix TDS1052B oscilloscope. A resistance box was used to analyze the output of the devices depending on the impedance load of the circuit. Different mechanical activations were produced, both manually pressing or bending or handwriting, and by machine, using the airflow produced by a fan or pressing and bending the devices with controlled movement of a magnetic shaker (Smart Shaker K2007E01 from The Modal Shop). The force applied by the magnetic shaker in pressing activations was measured with a force sensor (IEPE model 1053V2 from Dytran Instrument, Inc.) attached to the shaker and covered with a PDMS foil (thickness lower than 2 mm).

**Coarse-grained Monte Carlo simulations**

To assist in the analysis of the experimental results, we have made use of kinetic Monte Carlo simulations based on a coarse-grained model developed to simulate the growth of anisotropic crystalline materials (such as ZnO) on surfaces with arbitrary geometries. A detailed description of the model can be found in reference [39]. Here we include a few of the most relevant features: 1) in the first place, the system is defined by a simple 3D cubic lattice with periodic boundary conditions along the directions of the growth plane. 2) The lattice is characterized by an integer array in which lateral overhangs and vacancies are allowed. 3) The material deposited onto the surface is simulated as a flux of arriving particles that mimics typical conditions encountered in PECVD: the trajectory followed by the particles toward the surface is characterized by azimuthal and polar angles, which are obtained from an angular distribution that follows a Maxwell-type distribution for the particle energies. Once the particle reaches the surface, there is a certain possibility (weighted by a sticking coefficient $S_0$) that it bounces and continues its path with another trajectory until it reaches a new site or until it reaches the roof of the simulation domain. In that case, the particle disappears and another particle with a different trajectory is generated. 4) In addition to deposition, the model includes surface diffusion via a single-site random walk mechanism, characterized by an Arrhenius-type jump rate, $\Gamma_l = (6D_0/a_0^2)exp(-\Delta E_l/KT)$. The pre-factor $6D_0/a_0^2$ define the time scale of the jump, with $D_0$ being a parameter of the model, which is interpreted as a microscopically averaged diffusion coefficient, and $a_0$ is the size of the particle. The activation energy, $\Delta E_l$, for a diffusion event is calculated in a simple bond-counting scheme, $\Delta E_l = nE_b + mE_W + \delta E$, where $E_b$ is the bond strength between particles of the film and $E_W$ between the film and the substrate. Finally, $\delta E$ is an extra energy term that accounts for the structural features associated with the specific material being simulated (see details in [39]). For wurtzite, the parameters of this extra term include an anisotropy ratio, $A_r$, that allows adjusting the growth rate of the material in the directions parallel and perpendicular to the c-axis. For the results shown in this paper, the value of $A_r$ was set at 0.75, which correspond to a growth velocity that is slightly larger perpendicular to c than along this axis. This choice permits the reproduction of most features observed in the real samples.

## Results and discussion

ZnO thin films were fabricated by PECVD on paper and on the gold electrodes previously deposited on paper as detailed in the Experimental Methods and Scheme 1. Figure 1 a) compares the XRD patterns obtained in the Bragg-Brentano configuration for the three systems, namely, paper reference support (red curve), ZnO deposited directly on paper (blue) and on the gold layer (black). Contrary to the standard (002) texturization, ZnO crystalline structure develops through several planes under these experimental conditions. The XRD patterns present peaks corresponding to (100), (101), (002), and also with a lower intensity, to (110), (103), (112), and (201). In the case of the ZnO film deposited on Au-decorated paper, the peak corresponding to (101) plane shows an enhanced intensity, addressing a higher texturization along such a plane for this sample (see the texture coefficients calculated from the XRD patterns following the method in ref. [30] in Figure S2 at the Supporting Information Section).

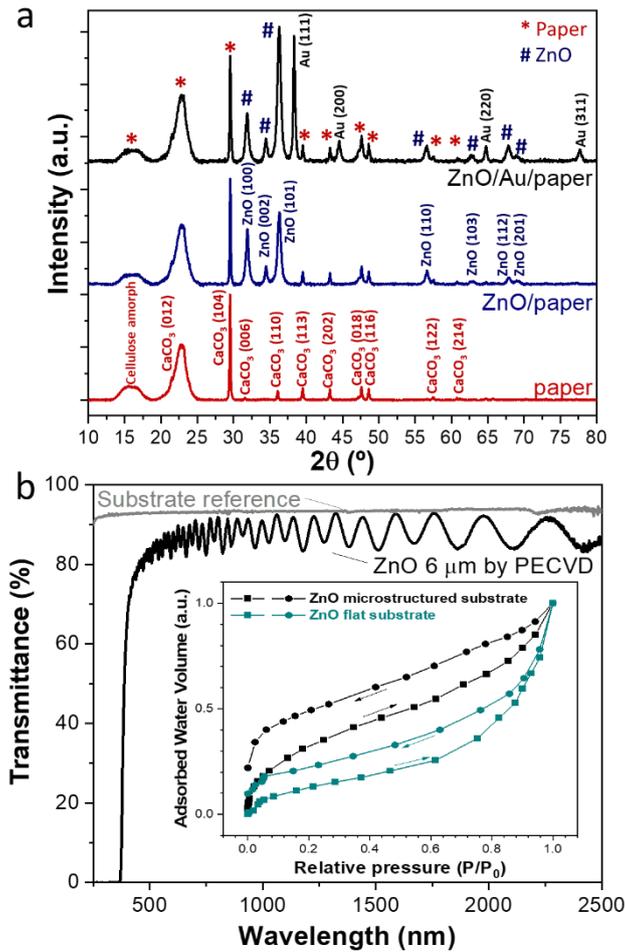

**Figure 1. Crystalline structure of the piezoelectric system, optical transparency, and porosity.** a) XRD diffraction patterns corresponding to ZnO thin films deposited on paper and gold-decorated paper. The main planes for polycrystalline Au (electrode), ZnO wurtzite (piezoelectric layer), amorphous cellulose, and $CaCO_3$ (commercially available paper) are labelled. b) UV-VIS-NIR transmittance spectrum for an equivalent 6 µm ZnO films deposited on fused silica showing high transparency above the ZnO bandgap wavelength. Inset in b) shows Adsorption/desorption isotherms of water on ZnO thin films prepared on the QCM flat substrate (cyan) and after Au particles decoration (black). The arrows indicate the adsorption (pointing right) and desorption (left) branches of the isotherms.

Figure 1 b) shows the UV-Vis-NIR transmittance spectra obtained for an equivalent ZnO thin film (black curve) deposited on a transparent fused silica substrate (grey). Thus, this coating is fully transparent (see also pictures in Figure S1) for a thickness close to 6 µm and presents a proper adhesion to the substrate. Such a feature is one of the great advantages of the application of PECVD instead of the standardized application of magnetron sputtering for the fabrication of ZnO piezoelectric layers, which normally involves additional thermal treatments to avoid thick-film related stress issues.[40,41]

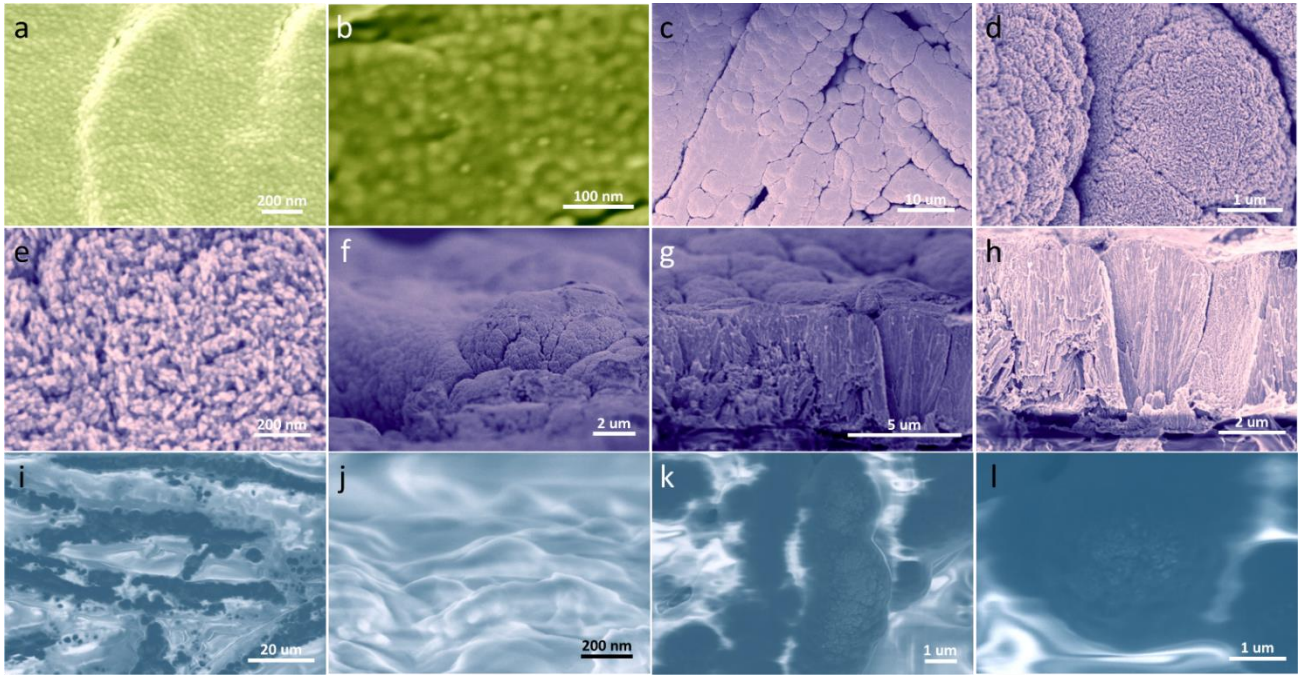

**Figure 2. Morphology and microstructure of the multilayer Au/ZnO/PMMA system on commercial paper substrates.** Characteristic SEM micrographs of the different synthetic steps for the materials directly deposited on paper in the vertically contacted device: Au layer formed by thermal evaporation (a,b); ZnO fabricated by PECVD on the gold interlayer (normal view c-e), tilted view (f) and cross-section (g-h); and spin-coated PMMA (tilted (I,j) and normal (k,l) views).

Thus, by PECVD in the downstream configuration, the fabrication of transparent crystalline ZnO is straightforwardly carried out at room temperature without the need for previous substrate treatments, temperature conditioning, or further annealing processes, which are procedures mostly incompatible with the use of paper substrates. This ensures the high growth rates deposition of low internal stress ZnO coatings deposited at room temperature on either rigid substrates (e.g. σ= -310 MPa and $\varepsilon_{zz}$ = 0.13 % for ZnO film deposited on fused silica) or temperature sensitive and/or flexible substrates (e.g. σ= 60 MPa and $\varepsilon_{zz}$ = -0.03 % for ZnO the coating deposited cellulose paper). These low stress and lattice-strain values, in comparison with other ZnO thin films deposited by RF sputtering at RT or low temperatures,[40–42] could be related to both the low ion bombardment during the PECVD process and the columnar porous structure,[35,43] of the films as analyzed below. The high stability of our PECVD films grown at RT is illustrated by the lack of interfacial delamination, buckling and or crack signs in the case of 6-μm thick films deposited at room temperature either on rigid or flexible substrates even after severe bending.

Figure 2 gathers the characteristic SEM micrographs for the different synthetic and processing steps. Please consider that depending on the final assembly, either vertical or lateral, the deposition of ZnO is carried out directly on the substrate or on the Au-decorated support (see Scheme 1). Firstly, a layer of Au is deposited by thermal evaporation with a thickness of ca. 100 nm (Fig. 2 a-b). Such a thickness allows for electrical connectivity providing bottom electrodes with a resistance below 20 Ω/sq. The surface of the Au layer is not completely flat, as it shows a two-scale roughness, in the microscale as a reflection of the paper fibres and the nanoscale due to the granular form of the gold. Similarly, the ZnO thin film also grows developing a multi-scale roughness by replication of the paper fibres (Fig. 2 c) in tens of microns and showing the characteristic crystalline microstructure with nano-scale grains randomly oriented (Fig. 2 e). It is worth mentioning, that at the mesoscale (Fig. 2 d), the morphology of the thin film presents interesting features, as globular formations at the skullcaps of the covered fibres (Fig. 2 d, f) and hand-fan-like cross-sectional sections (Fig. 2 g-h) (we will further discuss such characteristic formations in Figure 3). In the next step, a PMMA is spin-coated on the ZnO producing a film covering the entire system (Fig. 2 i), apparently also the hills produced by the ZnO conformal deposition on the paper fibres (Fig. 2 j). Finally, in the case of vertically contacted devices, a top contact electrode is formed by the thermal evaporation of Au as detailed in the experimental section. The sheet resistance of the top contact gold electrode is below 13 Ω/sq as characterized by the four-probe tester. SEM micrographs in Fig. 2 show that ZnO thin films growing on the microstructured and rough substrates develop a complex morphology in which besides the expected microporosity characteristic of PECVD polycrystalline films (porous diameter below 2 nm), mesopores (2 – 50 nm) and macroporous (higher than 50 nm) can be revealed. Although traditionally compact and single-crystalline piezoelectric devices (sensors and nanogenerators) have prevailed in the literature, there is a recent trend to produce controlled porous systems to enhance the surface-related applications of these nanomaterials. Concretely, it has been reported the positive effect of mesopores appearing at the interface of the ZnO thin films deposited by magnetron sputtering with the silicon substrate after high-temperature annealing (up to 950 ºC) for the development of piezoelectric sensors and nanogenerators.[44] Kim et al. showed the enhanced piezoelectric performance of porous P(VDF-TrFE) electrospun nanofibers produced under ambient humidity-induced phase separation. The porous nanofiber energy harvesters outperformed the equivalent smooth counterparts in voltage and power output even though the compact system was richer in the most electrically active β phase.[45] Also, there has been an effort in the implementation of porous piezoelectric and ferroelectric materials in piezo-catalysis applications devoted, for instance, to water splitting and piezo-photodegradation of water and air pollutants.[46] Aiming to corroborate the presence of such porosity in the ZnO thin films deposited by PECVD, we carried out room-temperature water adsorption isotherms by QCM (see details on the method elsewhere [37,38]). It must be stressed that for such a characterization, the ZnO thin films were deposited directly on the QCM crystal, with a much lower roughness than the cellulose. For the sake of a fair comparison, we decorated the QCM chips with Au nanoparticles with diameters in the order of 100 – 200 nm. The obtained isotherm cycles appear in the inset of Fig. 1 b). Both curves are representative of porosity including micro and mesopores (please note that macroporosity is not accessible by this characterization protocol), depicting the curve corresponding to the ZnO deposited on Au particles a higher mesoporosity (wider hysteresis loop) and also an increment of the microporous (higher cut with the y-axis).[37,38]

To shed light on the mechanisms behind such a complex microstructure and porous network development, we analyze the growth of ZnO by PECVD applying Coarse-grained Monte Carlo simulations. Details on this methodology can be found in a recent reference[39] where we presented the growth of polycrystalline ZnO thin films on flat substrates. In Figure 3 we show the effect of adding submicrometric regular features on the ZnO microstructure (morphology and texture) development. It is worth stressing the evident matching between the SEM cross-sectional micrographs obtained for the ZnO grown on paper decorated with Au (Fig. 3 left) and the simulations (Fig. 3 right). The growth mechanism in the first approach is dominated by the interplay between the shadowing effect imposed by the Au particles and the anisotropy of the growth velocity with respect to local crystallographic orientation. Thus, over the Au particles, the PECVD ZnO grows in fan-like mode and develops a highly intricating and fibre-like textured nanostructure characterized by nano and micropores which appear open in a great majority. It is easily distinguishable also a higher porosity rising from the intersections of ZnO growing from adjacent Au particles. Fig. 3 h, k, l) present the results of the Coarse-grained Monte Carlo simulations for the textural growth of the ZnO (and g, i, j the same results without textural colours for a better comparison with the SEM pictures). The visible angular distribution of the fan-like growth is also affecting the orientation regarding the c-axis as depicted by the change in the colour scale from grey (c-axis parallel to the substrate), to dark blue/red (c-axis perpendicular to the substrate as indicated in the colour map in Fig. 3 f). Figure 3 l) shows that for fan-like nanostructures grown on Au NPs, the c-axis may sustain high tilt angles with the vertical direction or even lay parallel to the substrate. The development of such highly tilted wurtzite nanostructures accounts for the higher intensity of the (101) peak of the samples deposited over the Au decorated substrates (cf. Fig. 1 a). In particular, the tilt angle θ = 60-80º calculated for the light blue areas in Fig. 3 k and l) agrees with the angle between the (001) and (101) planes in the wurtzite crystal.[47,48] On the contrary, in the boundaries between two adjacent fans, the c-axis appears with a preferent texture perpendicular to the substrate. In the case of the well-defined fan structure depicted in Figure 3 i, k) the texture distribution is homogeneous up to the surface, where patches of different colours (textures) appear. Moreover, the preferential texturization also depends on the distance between features on the substrate, as seen from the comparison of panels k) and l), showing that for closer particles the fan-like growth is less dominant in good agreement with a weaker shadowing effect. These results are in concordance with the XRD patterns in Fig. 1 a) and texturization coefficients gathered in Figure S2. It is important to note that such distribution of textures can make the system piezoactive under activation from different directions, as we will discuss below.

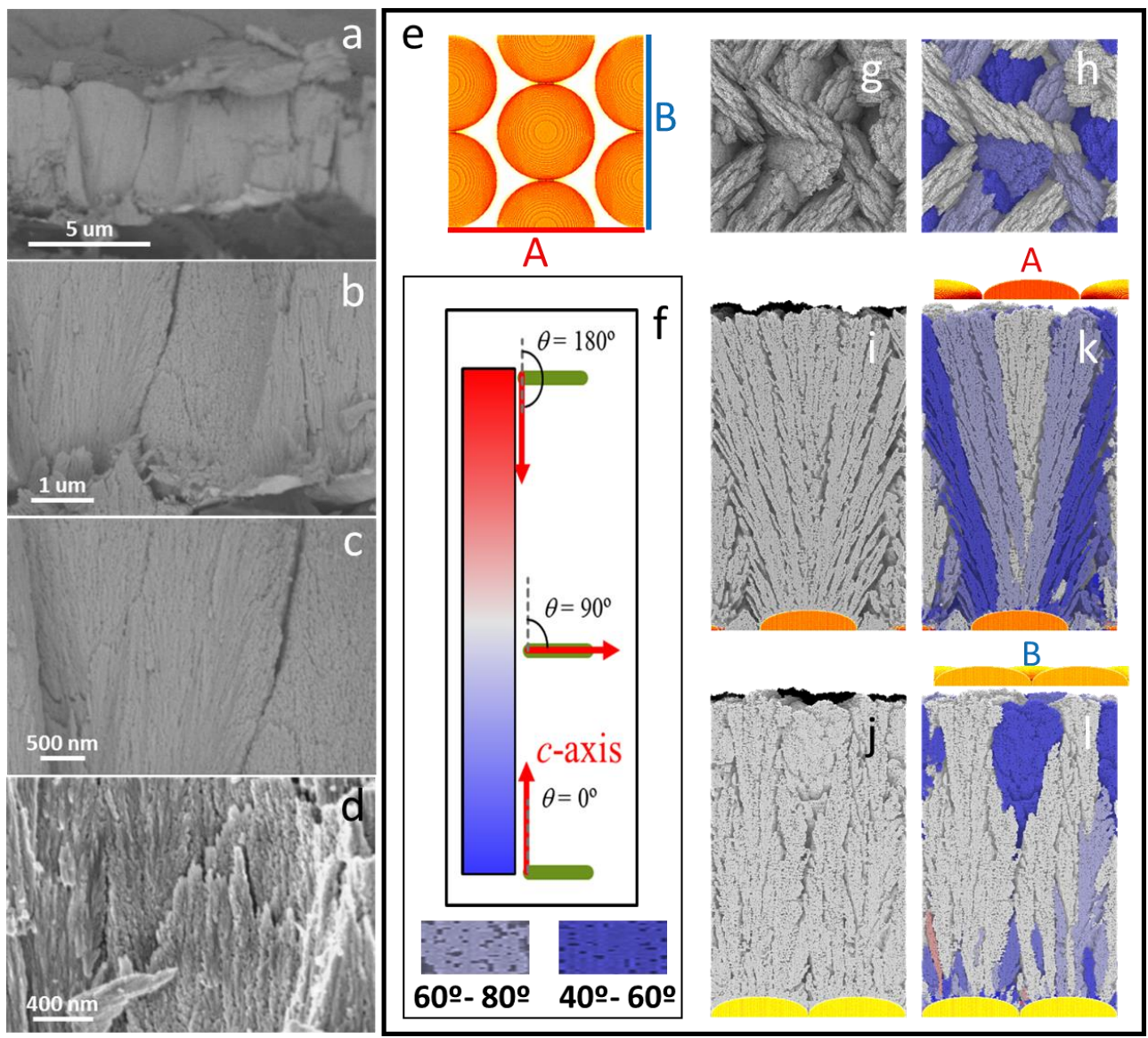

**Figure 3. Coarse-grained Monte Carlo simulations on the nanostructured and porous ZnO.** Left) cross-section SEM micrographs for backscattered (a-c) and secondary (d) electrons taken at different magnifications revealing the presence of micro and mesopores. Right) Top (g-h) and cross-sectional (i-l) views of the simulation corresponding to the material growth and porous development (g, i, j) and the crystalline arrangement (h, k, l) attending to the distance between particles at the interface with the substrate (distances labelled as A and B in figure e) and crystalline orientation (represented as the angle with respect to the c-axis, figure f).

**Figure 4** presents an overview of the characterization and application of the laterally contacted system as PENG. This architecture has been previously approached by other authors working with hydrothermally grown ZnO nanowires[49] and successfully tested as touch sensors[50] and with branched ZnO nanotrees.[51] In Fig. 4, the performance of the plasma-produced devices was characterized by following the short circuit current signal and output power through different loads under different actuation modes. Panel a) shows the response upon actuation with a magnetic shaker for fixed force and frequency as labelled. It is worth stressing the high repeatability of the outcome signal and the fast response of the system upon frequencies as high as 10 Hz (see also Figure S3). The power vs load curve in Fig. 4 b) depicts the characteristic peaky curve for piezoelectric nanogenerators with maxima at impedance values ca. $10^8$ Ω. It is also important to highlight the durability of the device. Thus, Fig. 4 c) compares the performance of the system after a lifetime longer than 10000 cycles of impact by the magnetic shaker. Such a feature is outstanding in comparison with the durability previously reported for ZnO NWs with complete degradation of the output current after 2000 cycles.[50] As is shown, the nanogenerators continue working with high repeatability and the total output current has only decreased by less than a 15 %.

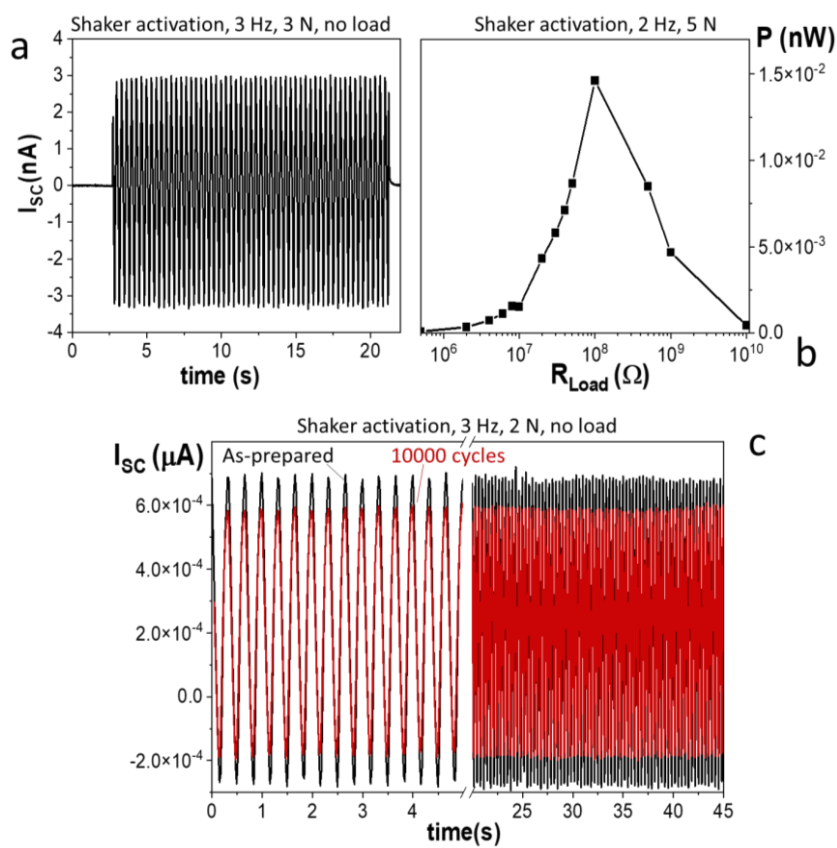

**Figure 4. Characteristic I-t and Power-Load curves and durability test for the laterally connected device.** a) Short-circuited current obtained under magnetic shaker tapping actuation for a constant force of 3 N and fixed frequency at 3 Hz; b) Power-load curve taken at 2 Hz and 5 N. c) c) Comparison of the Isc for laterally connected device as-prepared (black curve) and after impacting for more than 10000 cycles (red curve).

The high value for the optimum load in Fig. 4 b) might hamper the direct use of the nanogenerators as a power source for low-power electronics but opens the path toward its implementation as a self-powered force and/or strain sensor.[52] Consequently, we tested the response of the devices under different stimuli, paying special attention to the configuration as a force sensor at low frequencies and forces, i.e. under conditions compatible with their use as a wearable sensor (see Figure 5). The responses were highly repetitive and reproducible with output signal values in good agreement with previously reported sensors relying on single-crystalline materials.[49,50,53,54] Interestingly, the load circuit conditioned the response of the sensor in a great manner. Thus, depending on the value of the impedance load, i.e. the resistance connected between the sensor electrodes, the force signal can be analyzed in two different ways: i) output signal is proportional to the force for high impedance loads; ii) output signal becomes proportional to the time derivative of the force for low impedance loads as in the short circuit current measurements. This effect can be easily observed under 500 and 10 MΩ, where the output voltage is in phase and shifted with respect to the applied force, as shown in Fig. 5 a and b), respectively. Moreover, in the low-impedance case characterizing the output short-circuit current, the signal is 90º shifted with respect to the applied sinusoidal force (Fig. 5 c). To facilitate the comparison, the time derivative of the applied force has been added to the graph in Fig. 5 d) for a constantly decreasing force, showing the matching with the short-circuit current.

Although the description of the physical effects behind such a phenomenological finding will be the aim of a forthcoming article, we have included in Supporting Information Section 1 a first approach to solutions of the differential transport equation governing the response of the system for different load impedance (see also reference [55]). Specifically, for the most likely situation, where the impedance load corresponds to an intermediate value (see also Fig. 5 b) and applying a sinusoidal activation $\boldsymbol{F \propto \sin(\omega t)}$, where $\boldsymbol{\omega}$ is the applied angular frequency, the output signal would be proportional to a linear combination of both sine (proportional to force) and cosine (proportional to the time derivative of the force) following the expression:

$$\boldsymbol{V \propto (\omega_0 \cos(\omega t) + \omega \sin(\omega t))} \quad \textbf{(Eq.2)}$$

where $\boldsymbol{\omega_0}$ is a parameter dependent on the geometry and dielectric constant of the material and proportional to $\boldsymbol{1/R}$, taking R as the impedance load.

The limit of detection for the self-powered sensor has been estimated at 0.1 N for a frequency as low as 0.5 Hz with an upper limit defined by our experimental set-up of 10 N and frequencies higher than 10 Hz (see also Figure S3). It is worth stressing that the development of piezoelectric force and strain sensors into deformable substrates, as is the case of paper, it is extremely appealing since it paves the way for the development of self-powered body motion sensors, which in this case, would profit from the differences orientation that the piezoelectric c-axis takes on the Au decorated samples (cf. Fig. 3).[56]

Literature reports on diverse strategies to improve the output power of ZnO PENGs, including the implementation of semiconducting and conducting polymers to reduce screening effects at the interface of the semiconducting piezoelectric material and the metal electrode; doping ZnO to increase the piezoelectric coefficient or modifying the PENG architecture to enhance the kinetic energy harvesting and favour electrical impedance matching.[2,57] Herein, we have explored a vertically contacted (top-bottom) electrode configuration to reduce the total resistance of the devices and enhance the power output. It must be noted that in this case, the role played by the PMMA interface is critical as it reduces not only pre-screening effects but also the highly probable short-circuiting issues that are likely to affect porous structures after prolonged tests.

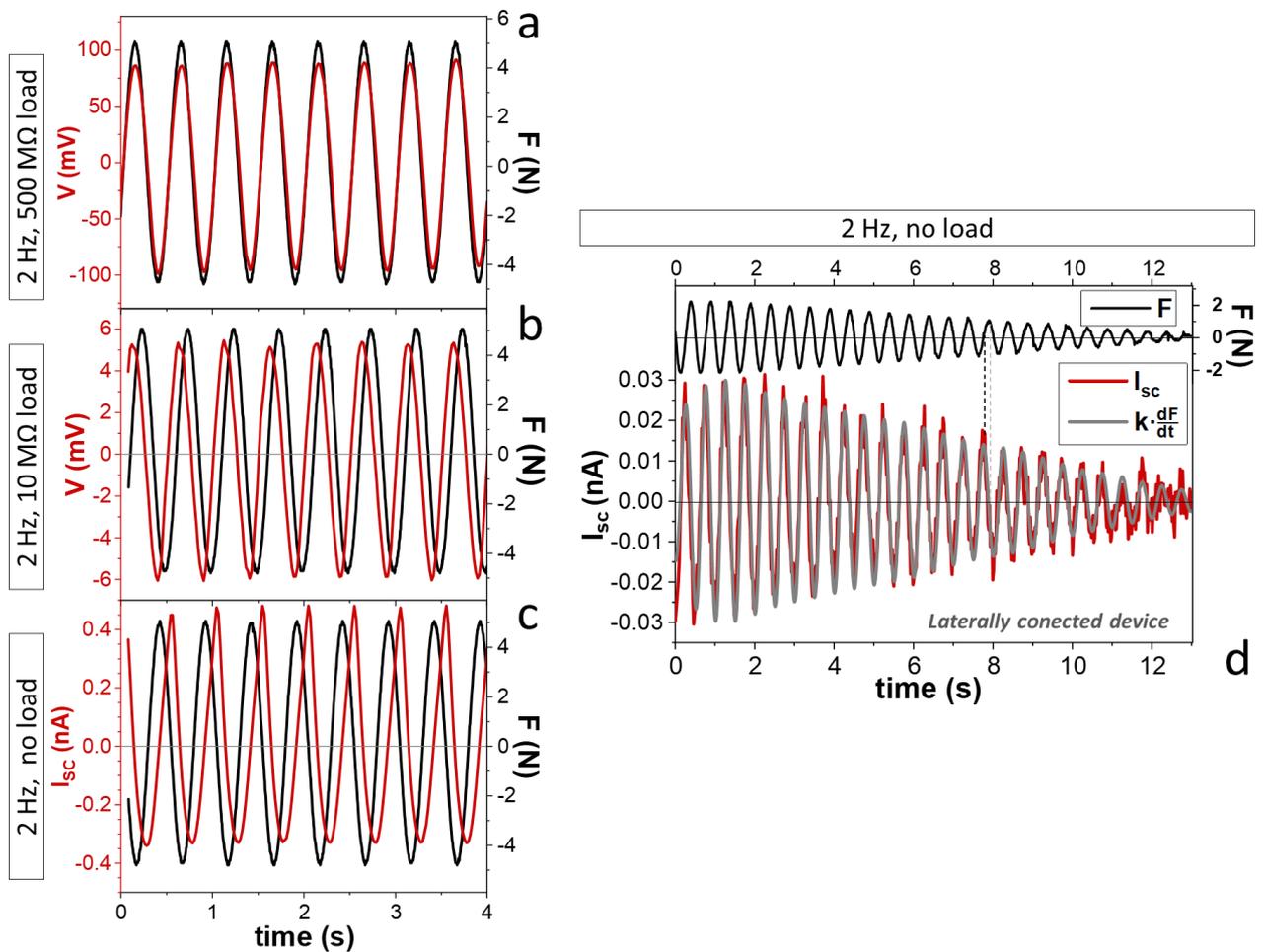

**Figure 5. Characteristic responses for the self-powered laterally connected sensor for three case studies of the load resistance.** a,b) V-t curves and c) I-t curves taken at 2Hz for a sinusoidal variable force (black) and different loads as indicated. d) I-t curve taken at 2Hz for a sinusoidal force of decreasing amplitude in short circuit configuration. The laterally connected sensor response is proportional to the time derivative of pressure force (grey curve) for low load impedances.

Figure 6 presents an overview of the characterization and application of the top-bottom system. Panel a) presents the I-V characteristic curve expected for an MIS Schottky heterojunction.[] Also, in comparison with the laterally contacted approach, the obtained output signal and power are much higher. Fig.6 b-d) gathers the output voltage and calculated power for increasing loads for the system vibrating at 25.7 Hz with a magnetic shaker assembled on a metallic cantilever (length of 15 cm). The maximum power appears for loads of ca. $3 \times 10^7\,\Omega$, as expected, an order of magnitude lower than the laterally connected counterpart. It is worth mentioning that the power delivered in this vertical configuration is much higher (4 orders of magnitude) than in the laterally connected architecture. Fig. 6 e-f) show the response to the same activation as the cantilever is manually bent from the short edge and suddenly released. It can be noted a higher voltage after the release of the cantilever, which is gradually attenuated as the vibration is dampened. We have also aimed to test the top-bottom connected device under real-case scenarios (Figure 7). First, these devices can be used as shape sensors taking advantage of the different characteristic curves when comparing different bending modes of the paper substrate. We can distinguish the position of the device on a plane (Fig. 7 a) or on a concave or curved surface or a corner (Fig. 7 b) by looking at the profile of the output current generated by low load mechanical excitation (manual tapping in this case). Thus, looking at the relative intensity values for the positive and negative peaks of the short-circuit current we get information about the sensor shape as the intensity of the positive/negative peaks is higher for the flat/concave configuration. Therefore, although the paper substrate could be folded at any angle and recover the original shape the signal generated by the device is determined by the particular arrangement of the paper when the mechanical stimulus is applied, providing a reliable response that can be used for positional sensing.

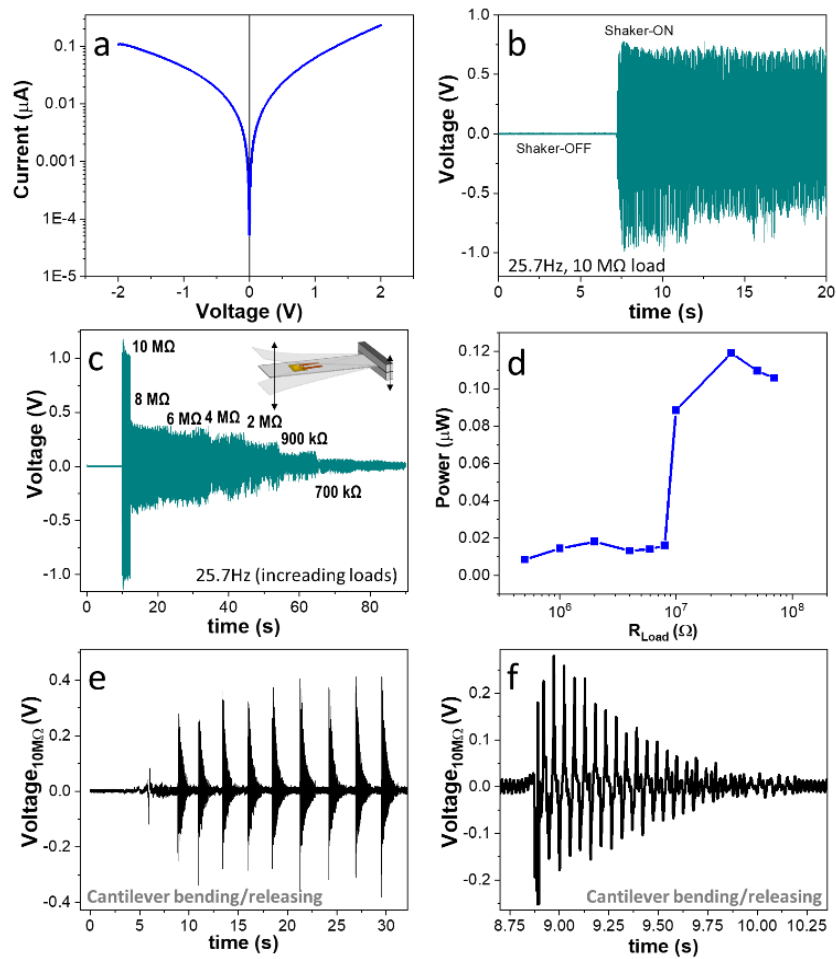

**Figure 6. Characteristic I-V and I-t curves for the vertically contacted PENG (cantilever mode).** a) I-V curve obtained for the PENG in cantilever mode after encapsulation in PMMA. b-c) Output voltage of a cantilever nanogenerator excited by a magnetic shaker at 25.7 Hz: for a constant resistance load of 10 MΩ (b) and for increasing loads (c). d) Measured power-load curve taken at 25.7 Hz; e) Response to the cantilever manual bending and release, curve in f) shows a zoom-in of the first cycle.

In the second real-case scenario, we exposed the devices to the airflow of a fan (Fig. 7 c-e). While the response of piezo and triboelectric nanogenerators to air flows have been described by several authors for the development of respiratory and airflow detection,[58,59] our interest here is to show the performance and robustness of the paper-based vertically contacted PENG under a continuous and intense mechanical stimulus. In the sequence of pictures of Fig. 7 e) (see also Video S1 as Supporting Information) it can be seen how the device is subjected to multiple random complex deformations. The intense airflow (aprox. 230 l/s) causes a constant combined bending, twisting, vibration and rotation of the device. The corresponding PENG signal is relatively complex as is the mechanical activation but the response is continuous for all spatial configurations.

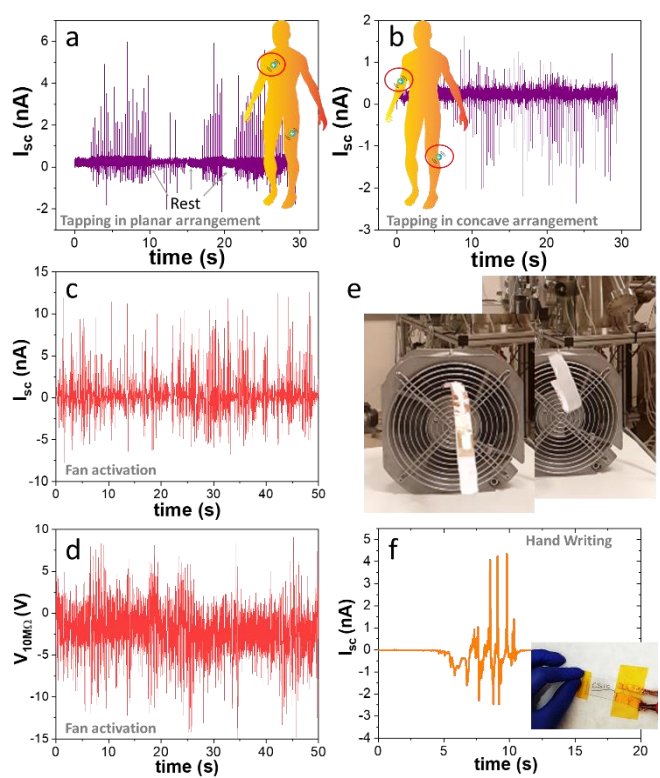

**Figure 7. Vertically contacted PENG responses for real-case scenarios.** a) Short circuit current obtained by finger tapping the top of the nanogenerator on a flat surface. b) Short circuit current obtained by finger tapping the top of the nanogenerator on a curved concave surface. Differences in a) and b) curves show that it could be possible to use the device as a shape sensor. c) Short circuit current and d) output voltage ($V_{10M\Omega}$) for the actuation of the vertical PENG with a fan as presented in the pictures in e). f) Example of the current generation by direct handwriting (ball pen) on the backside of the device.

In a first approximation, the signal resembles a combination of the signals obtained by mechanical activation of the flat and curved surface (Fig. 7 a-b) presenting an aperiodic distribution of positive and negative peaks of different intensities. Interestingly, the distribution of these peaks can be explained by the distribution of textures observed in the analysis and simulation of the ZnO piezoelectric layer proving the device can respond to mechanical activations from different directions (see Fig. 3 and the corresponding discussion). It is worth stressing that the weak point of this device is the Cu tape used as macro contact, which is often detached after such a mechanically extenuating experiment. However, the thin-film multilayer system (i.e. Au, ZnO and PMMA components) presents no cracks or delaminations, which is remarkable because of the large thickness of the piezoelectric film (up to 6 µm). As indicated above, such a good performance of these films fabricated by the PECVD plasma technique can be explained through the low structural and interfacial stress due to the low ion bombardment, in comparison with RF sputtering methods, and the porous and columnar structure. This scenario can be also improved thanks to the RT and solventless deposition on stretchable substrates which provides an additional route for stress relief.

Finally, the top-bottom devices also respond to the actuation by handwriting, as shown in Fig. 7 f). The output short-circuit current generated by direct handwriting with a ball pen on the backside of the device produces a specific signal pattern depending on the characteristics of the stroke such as the speed of execution and pressure in each of the text sections that can enable their application in anti-counterfeiting or handwriting recognition. This approach has been successfully implemented by different authors by producing arrays of self-powered sensors and studying the handwriting-generated signal by machine learning methods using relatively complex self-powered sensing architectures.[60–63] Besides, our approach is compatible with the reported handwriting analysis strategies but uses conventional paper, which permits the incorporation of the surface where the handwriting is naturally used. It is also worth mentioning that the reported self-powered writing recognition strategies are based almost entirely on triboelectric generators.[29,60] We expect that the use of a piezoelectric self-powered nanogenerator for the construction of writing recognition structures might add the advantage of using the pressure sensitivity characteristic of the deformation of piezoelectric structures providing richer information from specific handwriting patterns such as a signature.

## Conclusions

This work has demonstrated the custom-designed self-powered piezoelectric nanosensors and nanogenerators based on ZnO films (up to 6 um) fabricated by a scalable plasma PECVD technique on commercially available paper substrates without the need for any pretreatment or conditioning. The physicochemical characterization of the materials and systems together with the growth modelling by Monte Carlo simulations have revealed the microstructure, and porosity and cristallne texture of the layers. The simulations were carried out using a coarse-grained version of the Kinetic Monte Carlo method that allows modeling large-scale polycrystalline microstructures such as those developed in this work. PECVD allows for the formation of porous and thick polycrystalline layers at room temperature, in a solventless way. The layers feature outstanding mechanical properties and robustness, related to the low structural and interfacial stress as shown by the XRD analysis. These films offer a critical advantage in transducing/harvesting from random movements and mechanical inputs enabled by the fan-like orientation of the crystalline grains. Two different devices have been assembled, laterally contacted self-powered sensors and top-bottom cantilever-like nanogenerators. Both the developed architectures are highly flexible, foldable, and adaptable to any kind of deformation.

These results open an alternative path to fabricate flexible and durable paper-based self-powered sensors and piezoelectric nanogenerators. The nanogenerator produces electric energy in response to soft forces and low-frequency actuation compatible with human movements. The flexible piezoelectric generator harvests energy from irregular multifrequency and chaotic mechanical movements such as wind energy. Importantly, the fabrication protocol does not limit the area of the nanogenerators as PECVD is a straightforwardly scalable fabrication technique and extendable to the fabrication of doped metal oxides to enhance the piezoelectric coefficient or to apply in the deposition of transparent conductive oxides.

On the other hand, laterally connected devices can be used as force sensors for the transduction of human movements and similar low-frequency stimuli. Outstandingly, under sinusoidal excitation, the response is easily adjusted by controlling the load impedance. Thus as it is theoretically and experimentally analysed the device output can be proportional to either the applied force, the time derivate of the force or provide a mixed signal, as required.

The flexible, shapeable, and low-profile structure of the paper-based device allows for the fabrication and implementation of wearable body sensor networks. An intriguing aspect of the self-powered sensors is that they are highly dependent on both the type of mechanical activation and the spatial arrangement of the paper-based device when this activation occurs. This aspect is crucial for the development of positional/shape sensors that can provide information on the shape of the sensor when it experiences mechanical stress which is a critical aspect for the design and development of functional wearable sensors. Our devices also respond to processes in which the change of the sensor spatial conformation is continuously complex and random as has been shown in the activation with intense air flows. These results are particularly interesting for next-generation device development because they demonstrate the robustness of devices built on a relatively delicate substrate such as paper. This work also revealed how continuous and random conformational changes can also be tracked resulting in a high number of signals per second reflecting the specific conformations of the device at any given time. It is important to point out the strong potential of our materials systems for the development of functional devices on inexpensive and flexible substrates that are easily assimilated by the environment such as paper, and without compromising device functionalities and performance. This approach can solve one of the critical bottlenecks in the development of any distributed sensor technology such as cost per device and, especially, the environmental problems arising from the distribution of a large number of sensors in the environment without the possibility of direct traceability. The use of paper can also eliminate the need for waste collection and recycling post-operation. Finally, the use of paper-based supports is an important step in bringing the self-powered sensing technology to maturity.

## Author Contributions

XGC, FJA and AG carried out the experiments on fabrication and characterization. JB was responsible for the Monte Carlo simulations. JRSV, AB, KO and AnaB did the conceptualization. AnaB wrote the original draft of the article and supervised it. All the authors corrected and contributed to the final version of the manuscript.

## Conflicts of interest

There are no conflicts to declare

## Acknowledgments

We thank the projects PID2019-109603RA-I00, TED2021-130916B-I00 and PID2019-110430GB-C21 funded by MCIN/AEI/10.13039/501100011033 and by "ERDF (FEDER) A way of making Europe, Fondos NextgenerationEU and Plan de Recuperación, Transformación y Resiliencia", and the Consejería de Economía, Conocimiento, Empresas y Universidad de la Junta de Andalucía (PAIDI-2020 through project US-1381057). XGC thanks the FPU program through the FPU19/01864 grant number and FJA to the EMERGIA Junta de Andalucia program. K.O. thanks the Australian Research Council and QUT Centre for Materials Science for partial support. The project leading to this article has received funding from the EU H2020 program under grant agreement 851929 (ERC Starting Grant 3DScavengers).

# SUPPORTING INFORMATION SECTION

## Paper-based ZnO self-powered sensors and nanogenerators by plasma technology

Xabier García-Casas,[1] Francisco J. Aparicio,[1,2]* Jorge Budagosky,[1,2]* Ali Ghaffarinejad,[1, §] Noel Orozco,[1] Kostya (Ken) Ostrikov,[3] Juan R. Sánchez-Valencia,[1] Ángel Barranco[1] and Ana Borrás[1]*

[1]Nanotechnology on Surfaces and Plasma Lab, Materials Science Institute of Seville (CSIC-US), c/ Américo Vespucio 49, 41092, Seville, Spain.
[2] Departamento De Física Aplicada I, Escuela Politécnica Superior, Universidad De Sevilla, C/ Virgen De África 7, 41011, Seville, Spain.
[3] School of Chemistry and Physics and QUT Centre for Materials Science, Queensland University of Technology (QUT), Brisbane, QLD 4000, Australia.
§Current address: Sensors and Smart Systems Group, Institute of Engineering, Hanze University of Applied Sciences, 9747 AS Groningen, The Netherlands.

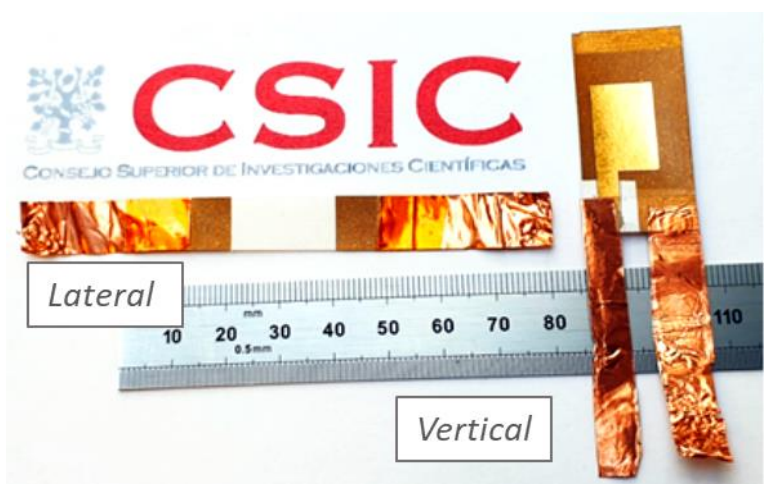

**Figure S1.** Picture showing examples of the two configurations, namely, laterally connected force-nanosensors and vertical (top-bottom) electrode cantilever energy harvester.

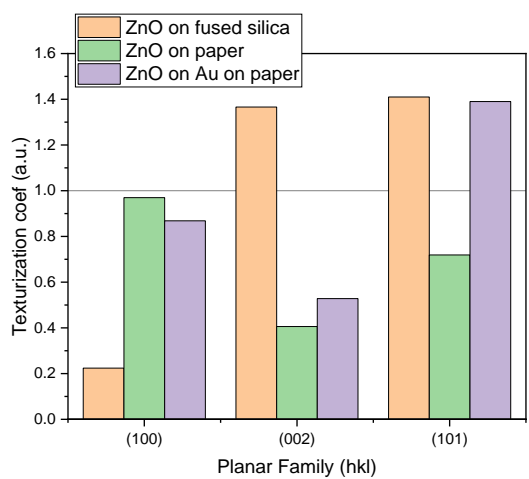

**Figure S2.** Texturization coefficients calculated from the Bragg-Brentano XRD patterns in Figure 1a) and applying the method gathered in ref. 1.

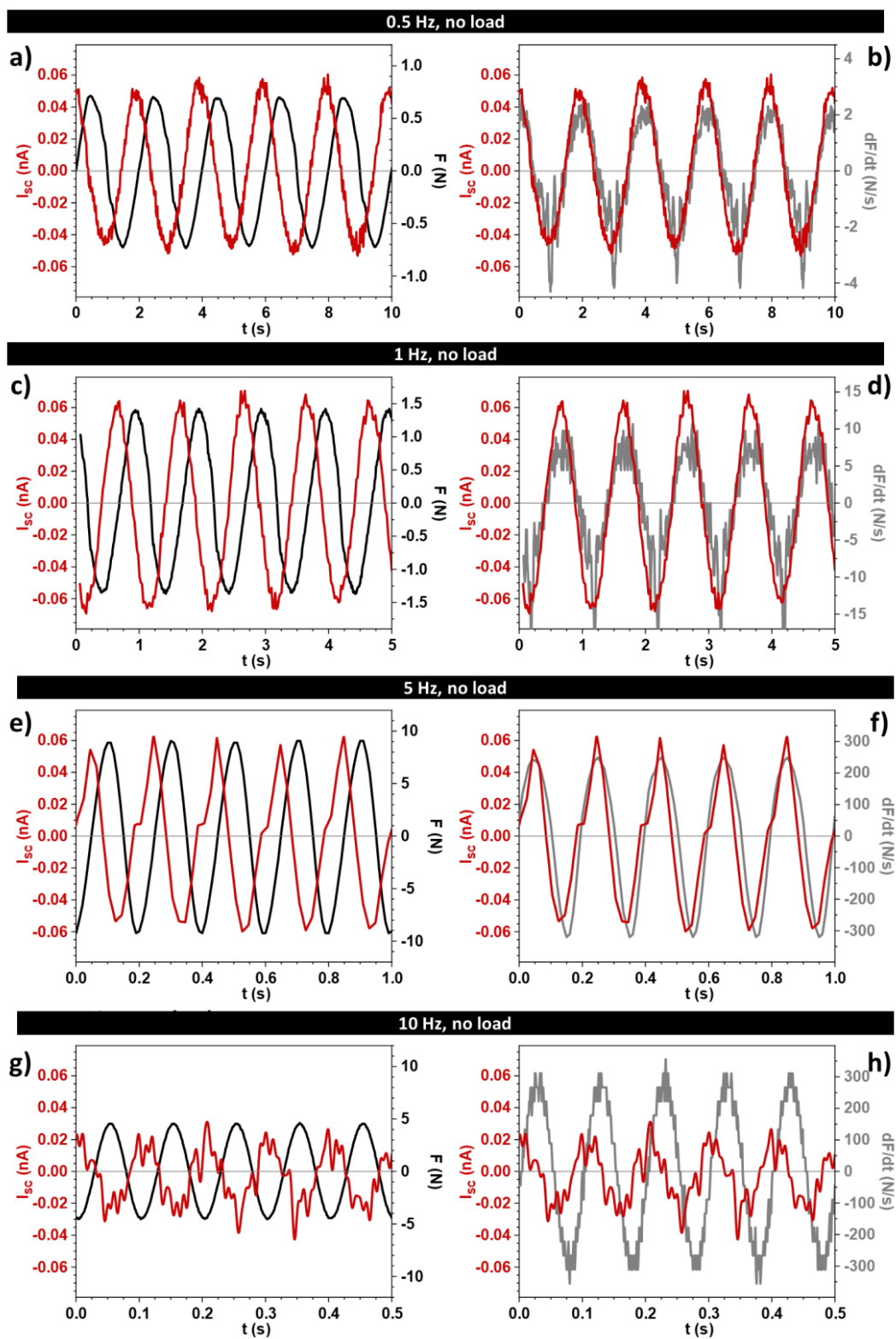

**Figure S3.** $I_{sc}$ – t curves for the laterally connected device as response to different frequencies (0.5, 1, 5 and 10 Hz) as labelled and for sinusoidal variable force. Comparison with the force (left panels) and with the derivative of the force with time (right panels).

**Video S1.** Paper-based nanogenerator moving under realistic conditions (fan stimuli). Available from the authors.

## Supporting Information Section 1: Theoretical derivation of the signal of piezoelectric nanogenerators under sinusoidal stimuli.

Given a piezoelectric nanogenerator of active area A, Load resistance R, and thickness z, the following equation describes the dependency on time of the induced charge density $\sigma(t)$ on its electrode [1]:

$$RA\frac{d\sigma}{dt} + z\frac{\sigma - \sigma_p(t)}{\varepsilon} = 0 \quad \textbf{(Eq. S1)}$$

where $\varepsilon$ is the electrical permittivity of the material and $\sigma_p(t)$ is the surface charge density produced by the polarization of the piezoelectric material under the stimulus. In this case, as the applied force is sinusoidal, then $\sigma_p(t) \propto \sin(\omega t)$

$$RA\frac{d\sigma}{dt} + z\frac{\sigma - k\sin(\omega t)}{\varepsilon} = 0 \quad \textbf{(Eq. S2)}$$

where $k$ is a proportionality constant related to the piezoelectric coefficient and the amplitude of the applied force. The homogeneous differential equation $RA\frac{d\sigma_H}{dt} + z\frac{\sigma_H}{\varepsilon} = 0$ has a general solution of the form:

$$\sigma_H = k_1 e^{-\omega_0 t} = k_1 e^{-\frac{z}{RA\varepsilon}t}$$

So that we get the relation between $\omega_0$ and the impedance load, the geometry of the device and the electrical permittivity. Taking $\sigma(t) = e^{-\omega_0 t} \cdot f(t)$ and substituting it again in Eq. S2

$$RA\frac{df}{dt}e^{-\omega_0 t} = z\frac{k\sin(\omega t)}{\varepsilon} \quad \textbf{(Eq. S3.1)}$$

$$\frac{df}{dt} = \frac{z}{RA\varepsilon}k\sin(\omega t)/e^{-\omega_0 t} = \omega_0 k e^{\omega_0 t}\sin(\omega t) \quad \textbf{(Eq. S3.2)}$$

Proceeding with integration by parts

$$\int e^{\omega_0 t}\sin(\omega t)dt = \frac{1}{\omega_0}e^{\omega_0 t}\sin(\omega t) - \frac{\omega}{\omega_0}\int e^{\omega_0 t}\cos(\omega t)\,dt$$

$$= \frac{1}{\omega_0}e^{\omega_0 t}\sin(\omega t) - \frac{\omega}{\omega_0}\left[\frac{1}{\omega_0}e^{\omega_0 t}\cos(\omega t) + \frac{\omega}{\omega_0}\int e^{\omega_0 t}\sin(\omega t)\,dt\right]$$

Thus

$$\left(1 + \frac{\omega^2}{\omega_0^2}\right) \cdot \int e^{\omega_0 t}\sin(\omega t)dt = \frac{1}{\omega_0}e^{\omega_0 t}\sin(\omega t) - \frac{\omega}{\omega_0}\frac{1}{\omega_0}e^{\omega_0 t}\cos(\omega t)$$

And we finally get

$$\int e^{\omega_0 t}\sin(\omega t)dt = \frac{\omega_0 e^{\omega_0 t}\sin(\omega t) - \omega e^{\omega_0 t}\cos(\omega t)}{\omega^2 + \omega_0^2}$$

We can substitute on Eq. S3.2 and get the general solution of Eq. S2 as

$$\sigma(t) = k_1 e^{-\omega_0 t} + \omega_0 k \frac{\omega_0 \sin(\omega t) - \omega \cos(\omega t)}{\omega^2 + \omega_0^2} \quad \textbf{(Eq. S4)}$$

As the output signal of the device is given by

$$V = RA\frac{d\sigma}{dt} \quad \textbf{(Eq. S5.1)}$$

$$I = V/R = A\frac{d\sigma}{dt} \quad \textbf{(Eq. S5.2)}$$

Therefore, for time much longer than $1/\omega_0$

$$V \propto I \propto (\omega_0 \cos(\omega t) + \omega \sin(\omega t)) \quad \textbf{(Eq. S6)}$$